\documentstyle[12pt,fleqn,paper,epsfig]{JHEP3}

\newcommand{\W}{{\vec W}}
\newcommand{\n}{\hat n}
\newcommand{\hn}{{\hat n}}
\newcommand{\hD}{{\hat D}}
\newcommand{\dfrac}{\displaystyle\frac}
\newcommand{\bea}{\begin{eqnarray}}
\newcommand{\eea}{\end{eqnarray}}
\newcommand{\oneg}{\displaystyle\frac{1}{g}}
\newcommand{\A}{{\vec A}}
\newcommand{\valpha}{{\vec \alpha}}

\newcommand{\D}{{\hat D}}
\newcommand{\nn}{\nonumber}
\newcommand{\pro}{\partial}

\title{\Large\bf Finite Energy Electroweak Dyon} \vskip 2em
\author{W. S. Bae\\C. N. Yang Institute for Theoretical Physics\\
State University of New York, Stony Brook, New York 11794, USA \\
E-mail:\email{wsbae@insti.physics.sunysb.edu}}
\author{Y. M. Cho\\C. N. Yang Institute for Theoretical Physics\\
State University of New York, Stony Brook, New York 11794, USA \\ 
and \\
School of Physics, College of Sciences, Seoul National
University \\ Seoul 151-742,
Korea\\E-mail:\email{ymcho@yongmin.snu.ac.kr}}

\abstract{ We present finite energy analytic monopole and dyon
solutions whose size is fixed by the electroweak scale. We discuss
two types of solutions. The first type is obtained by regularizing
the recent solutions of Cho and Maison by modifying the coupling
strength of the quartic self-interaction of $W$-boson in
Weinberg-Salam model. The second is obtained by enlarging the
gauge group $SU(2) \times U(1)$ to $SU(2) \times SU(2)$. Our
result demonstrates that one could actually construct genuine 
electroweak monopole and dyon
whose mass scale is much smaller than the grand
unification scale, with a minor modification of the electroweak
interaction without compromizing the underlying gauge invariance.}
\begin{document}
\setcounter{section}{1}
\setcounter{equation}{0}
\section*{I. Introduction}
\indent

Ever since Dirac \cite{Dirac} has introduced the concept of the
magnetic monopole, the monopoles have remained a fascinating subject
in theoretical physics.
The Abelian monopole has been generalized to the non-Abelian
monopoles by Wu and Yang \cite{Wu,cho80} who showed that
the pure $SU(2)$ gauge theory allows a point-like monopole,
and by 't Hooft and Polyakov \cite{Hooft,Julia} who have constructed
a finite energy monopole solution in Georgi-Glashow model
as a topological soliton.
In the interesting case of the electroweak theory of Weinberg and
Salam, however, it has generally been asserted that there
exists no topological monopole of physical interest \cite{vach}.
The basis for this ``non-existence theorem'' is, of course, that with
the spontaneous symmetry breaking the quotient space $SU(2) \times
U(1)/U(1)_{\rm em}$ allows no non-trivial second homotopy.
This has led many people to conclude that
there is no topological structure in Weinberg-Salam model 
which can accommodate a spherically symmetric magnetic monopole.

This, however, has been shown to be not true. Indeed some
time ago Cho and Maison~\cite{Cho0} have established
that Weinberg-Salam model and Georgi-Glashow model
have exactly the same topological
structure, and demonstrated the existence of a new type of
monopole and dyon solutions in the standard Weinberg-Salam
model.
{\em This was based on the observation that
the Weinberg-Salam model, with the hypercharge $U(1)$,
could be viewed as a gauged $CP^1$ model in which the (normalized)
Higgs doublet plays the role of the $CP^1$ field.} So
the Weinberg-Salam model does have exactly the same nontrivial
second homotopy as Georgi-Glashow model which allows topological monopoles.
This would have been impossible without the hypercharge $U(1)$.
Once this is understood, one could proceed to construct the desired monopole
and dyon solutions in Weinberg-Salam model.
Originally the solutions of Cho and Maison were
obtained by a numerical integration.
But a mathematically rigorous existence proof has
since been established which endorses
the numerical results, and the solutions are now referred
to as Cho-Maison monopole and dyon \cite{Yang1,Yang2}.

It should be emphasized that the Cho-Maison monopole is
completely different from the ``electroweak monopole''
derived from the Nambu's electroweak string. 
In his continued search for the string-like objects in physics,
Nambu has demonstrated the existence of a rotating dumb bell
made of the monopole anti-monopole pair connected by 
the neutral string of $Z$-field flux in Weinberg-Salam model \cite{nambu}.
Taking advantage of the Nambu's pioneering work,
others have claimed to discover an electroweak monopole,
simply by making the neutral string infinitely long and removing
the anti-monopole attached to the other end to infinity \cite{vacha}.
This type of ``electroweak monopole'', however, must carry a fractional
magnetic charge which can not be isolated, and 
obviously has no spherical symmetry which is manifest
in the Cho-Maison monopole \cite{Cho0}.

The Cho-Maison monopole may be viewed as a hybrid between the Abelian monopole
and the 't Hooft-Polyakov monopole, because it has a $U(1)$ point
singularity at the center even though the $SU(2)$ part is completely
regular. Consequently it carries an infinite energy
at the classical level, which means that physically
the mass of the monopole remains arbitrary.
{\em A priori} there is nothing
wrong with this, but nevertheless one
may wonder whether one can have an analytic
electroweak monopole which has a finite
energy. This has been shown to be possible \cite{cho97,cho02}.
The purpose of this paper is to discuss the finite energy
electroweak monopole and dyon solutions in detail.

The paper is organized as follows. In Section II we review the Cho-Maison 
monopole and dyon in Weinberg-Salam model, and discuss
the difference between the Cho-Maison monopole and the 
``monopole'' attached to Nambu's electroweak string.
In Section III we compare the Cho-Maison dyon with the Julia-Zee
dyon to clarify the similarities between the two dyons.
In doing so, we present the gauge-independent Abelian
formalism of the Weinberg-Salam model. In Section IV we 
show how a minor modification of the electroweak interaction
allows us, without compromizing the gauge invariance,  
to construct the finite energy electroweak monopoles 
and dyons. Utilizing the gauge-independent
Abelian formalism, we construct two types of solutions. 
The first is obtained by modifying the
coupling strength of the quartic self-interaction of $W$-boson.
The second is obtained by enlarging the gauge group to $SU(2) \times SU(2)$.
In both cases the gauge-independent Abelian formalism
plays a crucial role to guarantee the gauge invariance of 
the modified electroweak interactions. Finally in Section V we discuss 
the physical implications of the finite energy electroweak
monopoles.

\setcounter{section}{2}
\setcounter{equation}{0}
\section*{II. Cho-Maison Dyon in Weinberg-Salam Model}
\indent

Before we construct the finite energy monopole and dyon solutions
in electroweak theory we must understand how
one could obtain the infinite energy solutions first. So we will briefly
review the Cho-Maison solutions in Weinberg-Salam
model.
Let us start with the Lagrangian which
describes (the bosonic sector of) the standard Weinberg-Salam model
\begin{eqnarray}
{\cal L} &=& -|{\cal D}_{\mu} \phi|^2
 -\frac{\lambda}{2}(\phi^\dagger \phi
 -\frac{\mu^2}{\lambda} )^2
 -\frac{1}{4} \vec F_{\mu\nu}^2
 -\frac{1}{4}G_{\mu\nu}^2 ,
\label{lag1}
\end{eqnarray}
\begin{eqnarray}
{\cal D}_{\mu} \phi
&=& (\partial_{\mu} -i\frac{g}{2} \vec \tau \cdot \vec A_{\mu}
- i\frac{g'}{2}B_{\mu}) \phi
= (D_\mu - i\frac{g'}{2}B_{\mu}) \phi,\nonumber
\end{eqnarray}
where $\phi$ is the Higgs doublet,
$\vec F_{\mu\nu}$ and $G_{\mu\nu}$
are the gauge field strengths of $SU(2)$ and $U(1)$ with the
potentials $\vec A_{\mu}$ and $B_{\mu}$, and $g$ and $g'$
are the corresponding coupling constants.
Notice that $D_{\mu}$ describes the covariant derivative  of the $SU(2)$
subgroup only. From (\ref{lag1}) one has the following equations of motion
\begin{eqnarray}
{\cal D}^2 \phi
   &=&\lambda (\phi^{\dagger} \phi - \frac {\mu^2}{\lambda}) \phi,
\nonumber
\end{eqnarray}
\begin{eqnarray}
D_{\mu} \vec F_{\mu\nu}
   &=&-\vec j_{\nu}
    =i\frac{g}{2}\Big[\phi^{\dagger} \vec \tau ({\cal D}_{\nu} \phi)
      -({\cal D}_{\nu} \phi)^{\dagger} \vec \tau \phi \Big],
\label{eqm1}
\end{eqnarray}
\begin{eqnarray}
\partial_{\mu} G_{\mu\nu}
   &=&-k_{\nu}
    =i\frac{g'}{2}\Big[\phi^{\dagger} ({\cal D}_{\nu} \phi)
-({\cal D}_{\nu} \phi)^{\dagger}
\phi \Big].
\nonumber
\end{eqnarray}
Now we choose the following static spherically symmetric ansatz
\begin{eqnarray*}
\phi =\frac{1}{\sqrt{2}}\rho(r)\xi(\theta,\varphi),
\end{eqnarray*}
\begin{eqnarray*}
\xi=i\left(\begin{array}{cc}
         \sin (\theta/2)\,\, e^{-i\varphi}\\
       - \cos(\theta/2)
      \end{array} \right),
\hspace{5mm} \n = -\xi^{\dagger} \vec \tau ~\xi = \hat{r},
\end{eqnarray*}
\begin{eqnarray}
\vec A_{\mu}
&=& \frac{1}{g} A(r)\partial_{\mu}t~\hat{n}
   +\frac{1}{g}(f(r)-1) ~\hat{n} \times \partial_{\mu} \hat{n}
\label{ansatz1},
\end{eqnarray}
\begin{eqnarray}
B_{\mu} &=&\frac{1}{g'} B(r) \partial_{\mu}t 
           + \frac{1}{g'}(1-\cos\theta) \partial_{\mu} \varphi,
\nonumber
\end{eqnarray}
where $(t,r,\theta,\varphi)$ are the spherical coordinates.
Notice that the apparent string
singularity along the negative z-axis in $\xi$ and $B_{\mu}$ is a pure
gauge artifact which can easily be removed with a hypercharge $U(1)$
gauge transformation. Indeed
one can easily exociate the strings by making the hypercharge $U(1)$
bundle non-trivial \cite{Wu}. So {\it the above ansatz describes a most
general spherically symmetric ansatz of a $SU(2) \times U(1)$ dyon}.
Here we emphasize the importance of the non-trivial $U(1)$
degrees of freedom to
make the ansatz spherically symmetric. Without the extra $U(1)$ the
Higgs doublet does not allow a spherically symmetric ansatz.
This is because the spherical symmetry for
the gauge field involves the embedding
of the radial isotropy group $SO(2)$ into the gauge group
that requires the Higgs field to be
invariant under the $U(1)$ subgroup of $SU(2)$. This is possible
with a Higgs triplet,
but not with a Higgs doublet \cite{Forg}. In fact, in the
absence of the hypercharge $U(1)$ degrees of freedom, the above ansatz
describes the $SU(2)$ sphaleron which is not spherically
symmetric \cite{Dashen}. To see this, one might try to remove
the string in $\xi$ with the $U(1)$ subgroup of $SU(2)$.
But this $U(1)$ will necessarily change $\hat n$
and thus violate the spherical symmetry. This means that
there is no $SU(2)$ gauge transformation which can remove
the string in $\xi$ and at the same time keeps
the spherical symmetry intact. The situation changes with the
inclusion of the extra hypercharge
$U(1)$ in the standard model, which naturally
makes $\xi$ a $CP^1$ field \cite{Cho0}. This allows the spherical symmetry
for the Higgs doublet.

To understand the physical content of the ansatz we now
perform the following gauge transformation on (\ref{ansatz1})
\begin{eqnarray}
\xi \longrightarrow U \xi = \left(\begin{array}{cc}
0 \\ 1
\end{array} \right),
~~~~~U=i\left( \begin{array}{cc}
        \cos (\theta/2)& \sin(\theta/2)e^{-i\varphi} \\
        -\sin(\theta/2) e^{i\varphi} & \cos(\theta/2)
\end{array}
\right),
\label{gauge}
\end{eqnarray}
and find that in this unitary  gauge we have
\begin{eqnarray}
\hat n \longrightarrow \left(
\begin{array}{c} 0\\
0\\
1 \end{array}
\right), ~~~~~\vec A_\mu \longrightarrow
\frac{1}{g} \left(
\begin{array}{c}
-f(r)(\sin\varphi\partial_\mu\theta
    +\sin\theta\cos\varphi \partial_\mu\varphi) \\
f(r)(\cos\varphi\partial_\mu \theta
    -\sin\theta\sin\varphi\partial_\mu\varphi) \\
A(r)\partial_\mu t +(1-\cos\theta)\partial_\mu\varphi
\end{array} \right),
\label{unitary}
\end{eqnarray}
So introducing the electromagnetic potential $A_\mu^{\rm (em)}$ and the
neutral $Z$-boson potential $Z_\mu$ with the Weinberg angle $\theta_{\rm w}$
\begin{eqnarray}
\left( \begin{array}{cc}
A_\mu^{\rm (em)} \\  Z_{\mu}
\end{array} \right)
&=& \left(\begin{array}{cc}
\cos\theta_{\rm w} & \sin\theta_{\rm w}\\
-\sin\theta_{\rm w} & \cos\theta_{\rm w}
\end{array} \right)
\left( \begin{array}{cc}
B_{\mu} \\ A^3_{\mu}
\end{array} \right) \nonumber\\
&=& \frac{1}{\sqrt{g^2 + g'^2}}
\left(\begin{array}{cc}
g & g' \\ -g' & g
\end{array} \right)
\left( \begin{array}{cc}
B_{\mu} \\ A^3_{\mu}
\end{array} \right) , \label{wein}
\end{eqnarray}
we can express the ansatz (\ref{ansatz1}) by
\begin{eqnarray}
\rho &=& \rho(r) \nn\\
W_{\mu} &=& \dfrac{1}{\sqrt{2}}(A_\mu^1 + i A_\mu^2)
=\frac{i}{g}\frac{f(r)}{\sqrt2}e^{i\varphi}
      (\partial_\mu \theta +i \sin\theta \partial_\mu \varphi), \nn\\
A_{\mu}^{\rm (em)} &=& e \Big( \frac{1}{g^2}A(r) +
    \frac{1}{g'^2}B(r) \Big) \partial_{\mu}t 
 + \frac{1}{e}(1-\cos\theta) \partial_{\mu} \varphi,  \nonumber \\
Z_{\mu} &=& \frac{e}{gg'}(A(r)-B(r)) \partial_{\mu}t,
\label{ansatz2}
\end{eqnarray}
where $\rho$ and $W_\mu$ are the Higgs field and the $W$-boson,
and $e$ is the electric charge
\begin{eqnarray*}
e=\frac{gg'}{\sqrt{g^2+g'^2}}=g\sin\theta_{\rm w}=g'\cos\theta_{\rm w}.
\end{eqnarray*}
This clearly shows that the ansatz is
for the electromagnetic monopole and dyon.

The spherically symmetric ansatz (\ref{ansatz1}) reduces
the equations of motion to
\begin{eqnarray}
\ddot{f} -\frac{f^2-1}{r^2}f
             =\Big(\frac{g^2}{4}\rho^2 - A^2\Big)f, \nonumber
\end{eqnarray}
\begin{eqnarray}
\ddot{\rho}
 +\frac{2}{r} \dot{\rho}
 -\frac{f^2}{2r^2}\rho
 =-\frac{1}{4}(A-B)^2\rho
 +\dfrac {\lambda}{2}\Big(\rho^2
 -\frac{2\mu^2}{\lambda}\Big)\rho \nonumber,
\end{eqnarray}
\begin{eqnarray}
\ddot{A}
 +\frac{2}{r}\dot{A}
 -\frac{2f^2}{r^2}A
 =\frac{g^2}{4}\rho^2(A-B), \label{spher}
\end{eqnarray}
\begin{eqnarray}
\ddot{B}
 +\frac{2}{r} \dot{B}
 =-\frac{g'^2}{4} \rho^2 (A-B). \nonumber
\end{eqnarray}
Obviously this has a trivial solution
\bea
f=0,~~~~~\rho=\rho_0=\sqrt{2\mu^2/\lambda},~~~~~A=B=0,
\eea
which describes the point monopole in Weinberg-Salam model
\bea
A_\mu^{\rm (em)} = \dfrac{1}{e}(1- \cos \theta) \partial_\mu \varphi.
\eea
This monopole has two remarkable features. First,
this is not the Dirac's monopole.
It has the electric charge $4\pi/e$, not $2\pi/e$ \cite{Cho0}.
Secondly, this monople naturally admits a non-trivial
dressing of weak bosons. Indeed, with the non-trivial dressing,
the monopole becomes the Cho-Maison monopole and dyon.

To see this let us choose the following boundary condition
\bea
&&f(0)=1,~~~~~\rho(0)=0,~~~~~A(0)=0,~~~~~B(0)=b_0, \nn\\
&&f(\infty)=0,~~~~~\rho(\infty)=\rho_0,~~~~~A(\infty)=B(\infty)=A_0.
\label{bc0}
\eea
Then we can show that the equation (\ref{spher}) admits a family of
solutions labeled by the real parameter $A_0$ lying
in the range \cite{Cho0,Yang2}
\bea
0 \leq A_0 < \rm min ~\Big(e\rho_0,~\dfrac{g}{2}\rho_0\Big).
\label{boundA}
\eea
In this case all four functions $f(r),~\rho(r),~A(r)$, and
$B(r)$ must be positive for $r>0$, and $A(r)/g^2+B(r)/g'^2$
and $B(r)$ become increasing functions of $r$.
So we have $0 \leq b_0 \leq A_0$. Furthermore, we have
$B(r)\ge A(r)\ge 0$ for all range, and $B(r)$ must approach
to $A(r)$ with an exponential damping. Notice that,
with the experimental fact $\sin^2\theta_{\rm w}=0.2325$,
(\ref{boundA}) can be written as
$0 \leq A_0 < e\rho_0$.

\begin{figure}
\epsfysize=7cm
\centerline{\epsffile{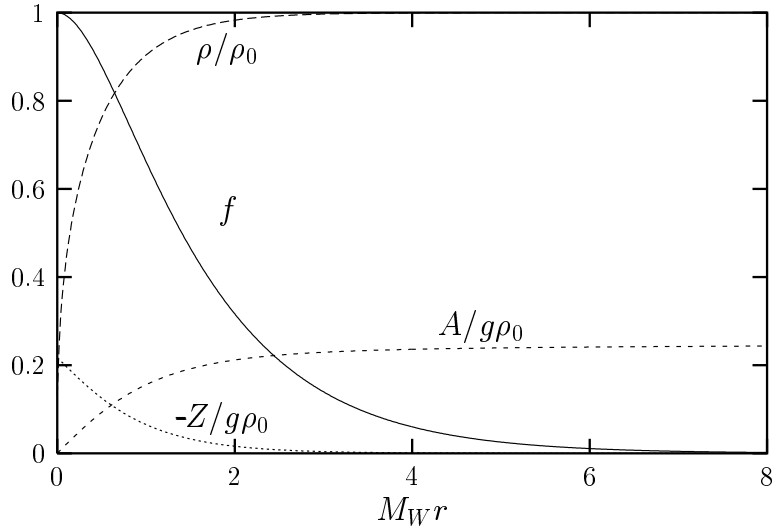}}
\caption{The Cho-Maison dyon solution. Here $Z(r)=A(r)-B(r)$
and we have chosen
$\sin^2\theta_{\rm w}=0.2325$, $\lambda/g^2=M_H^2/4M_W^2=1/2$,
and $A_0=M_W/2$.}
\label{fig1}
\end{figure}

With the boundary condition (\ref{bc0}) we can integrate
(\ref{spher}). For example, with $A_0=0$, we have the Cho-Maison monopole
with $A=B=0$. In general, with $A_0\ne0$, we find
the Cho-Maison dyon solution shown in Fig.\ref{fig1}~\cite{Cho0}.
The solution looks very much like the well-known
Prasad-Sommerfield solution of the Julia-Zee dyon.
But there is a crucial difference.
The Cho-Maison dyon now has a non-trivial
$A-B$, which represents the non-vanishing neutral $Z$-boson content
of the dyon as shown by (\ref{ansatz2}).

Near the origin the dyon solution
has the following behavior,
\begin{eqnarray}
\label{origin}
f    &\simeq& 1+ \alpha_1  r^2, \nonumber \\
\rho &\simeq& \beta_1 r^{\delta_-}, \nonumber \\
A    &\simeq& a_1 r,\\
B    &\simeq& b_0 + b_1 r^{2\delta_+} ,
          \nonumber
\end{eqnarray}
where $\delta_{\pm} =(\sqrt{3} \pm 1)/2$.
Asymptotically it has the following behavior,
\begin{eqnarray}
\label{infty}
f    &\simeq&  f_1 \exp(-\omega  r),\nonumber\\
\rho &\simeq& \rho_0 +\rho_1\frac{\exp(-\sqrt{2}\mu r)}{r}, \nonumber \\
A    &\simeq& A_0 +\frac{A_1}{r}, \\
B    &\simeq& A +B_1 \frac{\exp(-\nu r)}{r}, \nonumber
\end{eqnarray}
where $\omega=\sqrt{(g\rho_0)^2/4 -A_0^2}$,
and $\nu=\sqrt{(g^2 +g'^2)}\rho_0/2$.
The physical meaning of the asymptotic behavior
must be clear. Obviously $\rho$, $f$, and $A-B$ represent
the Higgs boson, $W$-boson, and $Z$-boson whose masses
are given by $M_H=\sqrt{2}\mu=\sqrt{\lambda}\rho_0$,
$M_W=g\rho_0/2$, and $M_Z=\sqrt{g^2+g'^2}\rho_0/2$. So
(\ref{infty}) tells that $M_H$, $\sqrt{1-(A_0/M_W)^2}~M_W$,
and $M_Z$ determine the exponential
damping of the Higgs boson, $W$-boson, and $Z$-boson
to their vacuum expectation values asymptotically.
Notice that it is $\sqrt{1-(A_0/M_W)^2}~M_W$, but not $M_W$,
which determines the exponential damping of the $W$-boson.
This tells that the electric potential of the dyon
slows down the exponential damping of the $W$-boson,
which is reasonable.

The dyon has the following electromagnetic charges
\begin{eqnarray}
\label{eq:Charge}
q_e &=& -{4\pi e}\bigg[
      r^2 \Big(\frac{1}{g^2}\dot{A}+\frac{1}{g'^2}\dot{B}
      \Big)\bigg] \bigg|_{r=\infty} =\frac{4\pi}{e} A_1
    =-\frac{8\pi}{e}\sin^2\theta_{\rm w}\int\limits^\infty_0 f^2 A dr , \\
q_m &=& \frac{4\pi}{e}. \nonumber
\end{eqnarray}
Also, the asymptotic condition (\ref{infty}) assures
that the dyon does not carry any neutral charge,
\begin{eqnarray}
\label{neutral}
Z_e &=&-\frac{4\pi e}{gg'}\Big[ r^2 (\dot{A}-\dot{B})\Big]
        \Big|_{r=\infty}
=0,\nonumber \\
Z_m &=& 0.
\end{eqnarray}
Furthermore, notice that the dyon equation (\ref{spher})
is invariant under the reflection
\bea
A \rightarrow -A,~~~~~~B\rightarrow -B.
\label{ref}
\eea
This means that, for a given magnetic charge,  
there are always two dyon solutions
which carry opposite electric charges $\pm q_e$.
Clearly the signature of
the electric charge of the dyon is determined by
the signature of the boundary value $A_0$.

With the ansatz (\ref{ansatz1})
we have the following energy of the dyon
\begin{eqnarray}
\label{eng1}
E=E_0 +E_1,
\nonumber
\end{eqnarray}
\begin{eqnarray}
E_0=\frac{2\pi}{g^2}\int\limits_0^\infty \frac{dr}{r^2}
\bigg\{\frac{g^2}{g'^2}
+ (f^2-1)^2\bigg\},
\nonumber
\end{eqnarray}
\begin{eqnarray}
E_1&=&\frac{4\pi}{g^2} \int\limits_0^\infty dr \bigg\{
\frac{g^2}{2}(r\dot\rho)^2
+\frac{g^2}{4} f^2\rho^2 +\frac{g^2r^2}{8} (B-A)^2 \rho^2
+\frac{\lambda g^2r^2}{8}\Big(\rho^2-\frac{2\mu^2}{\lambda}\Big)^2
\nonumber \\
&& +(\dot f)^2
+\frac{1}{2}(r\dot A)^2
+\frac{g^2}{2g'^2}(r\dot B)^2
+  f^2 A^2 \bigg\}.
\end{eqnarray}
The boundary condition (\ref{bc0}) guarantees that
$E_1$ is finite.
As for $E_0$ we can minimize it with the boundary condition $f(0)=1$, but
even with this $E_0$ becomes infinite.
Of course the origin of this
infinite energy is obvious,
which is precisely due to the magnetic singularity of $B_\mu$ at
the origin.
This means that one can not predict the mass of dyon.
Physically it remains arbitrary.

The numerical solutions assures the existence of
the electroweak monopole and dyon in  Weinberg-Salam model.
In spite of this one may
still like to have a mathematically rigorous existence proof
of the Cho-Maison solutions. The mathematical existence proof
is non-trivial, because the equation of motion (\ref{spher}) is not
the Euler-Lagrange equation of the positive definite
energy (\ref{eng1}), but that of the indefinite action
(\ref{lag1}). Fortunately the existence proof
has been given by Yang \cite{Yang1,Yang2}.

At this point it should be mentioned that the existence of 
a different type of ``electroweak monopole'' has been asserted 
in the litereature which has a fractional magnetic charge \cite{vacha}, 
\bea
\tilde q_m = \dfrac{4\pi}{e} \sin \theta_{\rm w}.
\eea 
This assertion has made a wrong impression that the fractionally
charged monopole is the only monopole which could exist 
in Weinberg-Salam model \cite{vach}, which has led many people 
to question the correctness
of the Cho-Maison monopole. So it is worth to clarify the 
situation before we close this section. 
As we have pointed out in the introduction, 
long time ago Nambu has shown the existence of electroweak 
string in Weinberg-Salam model which has monopole 
anti-monopole pair with the fractional charge 
$\pm \tilde q_m$ at the ends \cite{nambu}. Obviously
one could try to isolate the monopole at one end by extending 
the string to infinity. Simply by doing this some people
have claimed to discover the fractionally charged 
``electroweak monopole'' \cite{vacha}.
But clearly the Nambu's monopole can not be identified as
an electroweak monopole, because one can not really isolate it.
To do so one has to pump in an infinite energy. 
This means that the fractionally charged monopoles,
just like the quarks in QCD, can only be paired with 
the anti-monopoles to form confined objects which can not be 
isolated with finite energy \cite{nambu}. 
Indeed this confinement of the fractionally charged
monopoles in the electroweak theory was precisely 
the motivation of the Nambu's pioneering work.

Even if one neglects the confinement and simply considers the monopole
configuration with infinite string, one can not regard it as an
electroweak monopole. The reason is because along the 
string the Higgs field must vanish,   
so that asmptotically the Higgs field does not approach
its vacuum value in the vicinity of the string. 
This forbids us to identify the monopole as
an electroweak object.       

Nevertheless, as far as the Higgs doublet is concerned,
the ansatz (\ref{ansatz1}) is identical to Nambu's ansatz.
If so, one may wonder how Nambu did not discover the Cho-Maison
monopole. The reason is that he concentrated on the trivial sector
of the hypercharge $U(1)$ bundle because he was interested in 
the string configuration. As long as one stays in 
the trivial sector, of course, one can not remove the string
and must treat it as physical.
This is how Nambu constructed the electroweak string which confines
the monopole anti-monopole pair. 
Notice, however, one can always make the string disappear 
by making the $U(1)$ bundle non-trivial. In this case 
one can easily remove the string by a gauge transformation, and have 
a genuine isolated monopole which has the integer magnetic
charge $4\pi/e$ \cite{Cho0} . This is how 
the spherically symmetric Cho-Maison monopole has been constructed.
This clarifies the difference between the Nambu's monopole
and the Cho-Maison monopole, which emphasizes again the fact that 
the non-trivialty of the hypercharge $U(1)$
bundle is crucual for the Cho-Maison monopole.     

\setcounter{section}{3}
\setcounter{equation}{0}
\section*{III. Comparison with Julia-Zee Dyon}
\indent

At this stage one may ask whether there is any way to make
the energy of the Cho-Maison solutions finite.
A simple way to make the energy finite is to introduce
the gravitational interaction~\cite{Bais}.
But the gravitational interaction is not likely remove the singularity
at the origin, and one may still wonder if there is any way to
regularize the Cho-Maison solutions.
To answer this question it is important to understand that
the finite energy non-Abelian monopoles are really nothing but the
Abelian monopoles whose singularity at the origin is regularized
by the charged vector fields. This can best be demonstrated
by the t'Hooft-Polyakov monopole in Georgi-Glashow model \cite{cho97,klee}.
So in this section we discuss the gauge invariant
Abelian formalism of Georgi-Glashow model and review
how the charged vector field regularizes the Abelian
monopole singularity at the origin.

Consider Georgi-Glashow model
\begin{eqnarray}
\label{eq:Julia}
{\cal L}_{GG} =-\dfrac{1}{2}(D_\mu \vec \Phi)^2
-\dfrac{\lambda}{4}\left(\vec \Phi^2-\frac{\mu^2}{\lambda}\right)^2
-\dfrac{1}{4} \vec F_{\mu\nu}^2 ,
\end{eqnarray}
where $\vec \Phi$ is the Higgs triplet.
A best way to Abelianize it is
to start from the gauge-independent decomposition
of the $SU(2)$ gauge potential into the restricted binding potential
$\hat A_\mu$ and the gauge covariant 
valence potential $\vec W_\mu$ \cite{cho1,cho2},
which has recently
been referred to as Cho decomposition
or Cho-Faddeev-Niemi-Shabanov decomposition \cite{fadd,shab}.
Let
\begin{eqnarray}
\label{eq:Tri}
\vec \Phi = \rho \hat n,
\end{eqnarray}
and identify $\n$ to be the unit
isovector which selects the charge direction
in $SU(2)$ space. Then the Cho decomposition of
an arbitrary $SU(2)$ gauge potential is given by \cite{cho1,cho2}
\bea \label{chodecom}
& \vec{A}_\mu =A_\mu \n - \oneg \n\times\pro_\mu\n+\W_\mu
         = \hat A_\mu + \W_\mu,
~~~~~(A_\mu = \n\cdot \vec A_\mu,~~~ \hat{n}\cdot\vec{W}_\mu=0),
\eea
where $A_\mu$ is the ``electric'' potential.
Notice that the restricted potential $\hat A_\mu$ is precisely
the connection which leaves $\n$ invariant under parallel transport,
\bea
\D_\mu \n = \pro_\mu \n + g {\hat A}_\mu \times \n = 0.
\eea
Under the infinitesimal gauge transformation
\bea
\delta \n = - \vec \alpha \times \n,
~~~~~\delta \A_\mu = \oneg  D_\mu \vec \alpha,
\eea
one has
\bea
&&\delta A_\mu = \oneg \n \cdot \pro_\mu \valpha,
~~~~~\delta \hat A_\mu = \oneg \D_\mu \valpha,
~~~~~\delta \W_\mu = - \valpha \times \W_\mu  .
\label{gt1}
\eea
This tells that $\hat A_\mu$ by itself describes an $SU(2)$
connection which enjoys the full $SU(2)$ gauge degrees of
freedom. Furthermore the valence potential $\vec W_\mu$ forms a
gauge covariant vector field under the gauge transformation.
{\it But what is really remarkable is that the decomposition is
gauge independent. Once
$\hat n$ is chosen, the decomposition follows automatically,
regardless of the choice of gauge} \cite{cho1,cho2}.

Remarkably $\hat{A}_\mu$ retains all the essential
topological characteristics of the original non-Abelian potential.
First, $\hat{n}$ defines $\pi_2(S^2)$
which describes the non-Abelian monopoles \cite{Wu,cho80}.
Secondly, it characterizes
the Hopf invariant $\pi_3(S^2)\simeq\pi_3(S^3)$ which describes
the topologically distinct vacua \cite{bpst,cho79}.
Furthermore $\hat{A}_\mu$ has a dual
structure,
\begin{eqnarray}
& \hat{F}_{\mu\nu} = \partial_\mu \hat A_\nu-\partial_\nu \hat A_\mu
+ g \hat A_\mu \times \hat A_\nu = (F_{\mu\nu}+ H_{\mu\nu})\hat{n}\mbox{,}
\nonumber \\
& F_{\mu\nu} = \partial_\mu A_{\nu}-\partial_{\nu}A_\mu \mbox{,}
~~~~~H_{\mu\nu} = -\dfrac{1}{g} \hat{n}\cdot(\partial_\mu
\hat{n}\times\partial_\nu\hat{n})
= \partial_\mu \tilde C_\nu-\partial_\nu \tilde C_\mu,
\end{eqnarray}
where $\tilde C_\mu$ is the ``magnetic'' potential.
Notice that one can always introduce the magnetic
potential, since $H_{\mu \nu}$ forms a closed two-form
locally sectionwise \cite{cho1,cho2}.
In fact, replacing $\hn$ with a $CP^1$ field $\xi$ by
\bea
\hn=-\xi^{\dagger} \vec \tau ~\xi,
\eea
we have
\bea
\tilde C_\mu = \dfrac{2i}{g} \xi^{\dagger} \partial_\mu \xi,
~~~~~H_{\mu\nu} = \dfrac{2i}{g} (\partial_\mu \xi^{\dagger}
\partial_\nu \xi - \partial_\nu \xi^{\dagger} \partial_\mu \xi).
\eea
To see that $\tilde C_\mu$ does describe the monopole,
notice that with the ansatz (2.3) we have
\bea
\tilde C_\mu = \dfrac{1}{g} (1- \rm cos\theta) \partial_\mu \varphi.
\eea
This is nothing but the Abelian monopole potential, which
justifies $\tilde C_\mu$ as the magnetic potential.
The corresponding non-Abelian monopole potential is given by
\bea
\label{eq:vecC}
\vec C_\mu= -\dfrac{1}{g}\hat n \times \partial_\mu\hat n ,
\eea
in terms of which the magnetic field is expressed by
\bea
\vec H_{\mu\nu}=\partial_\mu \vec C_\nu-\partial_\nu \vec C_\mu+ g
\vec C_\mu \times \vec C_\nu = H_{\mu\nu}\hat n.
\eea
This provides the gauge independent separation of the monopole
field $\vec H_{\mu\nu}$ from the generic non-Abelian
gauge field $\vec F_{\mu\nu}$. The monopole potential (\ref{eq:vecC})
is now referred to as the Cho connection by Faddeev \cite{fadd,shab}.

With the decomposition (\ref{chodecom}), one has
\bea
\vec{F}_{\mu\nu}&=&\hat F_{\mu \nu} + \D _\mu \W_\nu -
\D_\nu \W_\mu + g\W_\mu \times \W_\nu,
\eea
so that the Yang-Mills Lagrangian is expressed as
\bea
&{\cal L}_{YM} =-\dfrac{1}{4}{\hat F}_{\mu\nu}^2 
-\dfrac{1}{4}(\D_\mu\W_\nu-\D_\nu\W_\mu)^2 
-\dfrac{g}{2}{\hat F}_{\mu\nu} \cdot (\W_\mu \times \W_\nu) \nn\\
&-\dfrac{g^2}{4} (\W_\mu \times \W_\nu)^2.
\eea
This shows that the Yang-Mills theory can be viewed as
a restricted gauge theory made of the restricted potential,
which has the valence gluons as its source \cite{cho1,cho2}.
For a long time it has generally been asserted that 
the non-Abelian gauge symmetry uniquely determines its dynamics.
This has led many people to believe that the Yang-Mills theory
is the only theory which has the full non-Abelian gauge symmetry.
But the above analysis clearly demonstrates that this is 
not true. {\it Evidently the non-Abelian gauge symmetry
allows a simpler gauge theory, the restricted gauge theory,
which enjoys the full non-Abelian gauge degrees of freedom
yet contains much less physical degrees of freedom}.
In this view the Yang-Mills theory is nothing but the
restricted gauge theory which has an extra gauge covariant
vector field as the colored source. This observation plays a central role 
in our understanding of QCD, in particular the monopole 
condensation in QCD \cite{cho1,cho2}. Only recently
this important fact has become to be appreciated \cite{fadd,shab}.

An important advantage of the decomposition (\ref{chodecom})
is that it can actually
Abelianize (or more precisely ``dualize'') the non-Abelian
gauge theory \cite{cho1,cho2}. To see this let
$(\hat n_1,~\hat n_2,~\hat n)$
be a right-handed orthonormal basis of $SU(2)$ space and let
\begin{eqnarray}
&\vec{W}_\mu =W^1_\mu ~\hat{n}_1 + W^2_\mu ~\hat{n}_2,
~~~~~(W^1_\mu = \hat {n}_1 \cdot \vec W_\mu,~~~W^2_\mu =
\hat {n}_2 \cdot \vec W_\mu).            \nonumber
\end{eqnarray}
With this one has
\begin{eqnarray}
&\hat{D}_\mu \vec{W}_\nu =\Big[\partial_\mu W^1_\nu-g
(A_\mu+ \tilde C_\mu)W^2_\nu \Big]\hat n_1
+ \Big[\partial_\mu W^2_\nu+ g (A_\mu+ \tilde C_\mu)W^1_\nu \Big]\hat{n}_2,
\end{eqnarray}
so that with
\bea
{\cal A}_\mu = A_\mu+ \tilde C_\mu,
~~~~~W_\mu = \dfrac{1}{\sqrt{2}} ( W^1_\mu + i W^2_\mu ), \nn
\eea
one could express the Lagrangian explicitly in terms of the dual
potential ${\cal A}_\mu$ and the complex vector field $W_\mu$,
\begin{eqnarray} \label{eq:Abelian}
{\cal L}_{YM} = -\dfrac{1}{4}({\cal F}_{\mu\nu}+ W_{\mu\nu})^2
-\dfrac{1}{2}|\hat{D}_\mu{W}_\nu-\hat{D}_\nu{W}_\mu|^2,
\end{eqnarray}
where now $\hat{D}_\mu=\partial_\mu + ig {\cal A}_\mu$
is an Abelian covariant  derivative, and
\bea
{\cal F}_{\mu\nu} = F_{\mu\nu} + H_{\mu\nu},
~~~~~W_{\mu\nu} = - i g ( W_\mu^* W_\nu - W_\nu^* W_\mu ).  \nonumber
\eea
This describes an Abelian gauge theory coupled to
the charged vector field $W_\mu$.
In this form the equations of motion of Yang-Mills theory is expressed by
\begin{eqnarray} \label{eq:AbelianEOM}
&\partial_\mu({\cal F}_{\mu\nu} +W_{\mu\nu}) = i g W^*_\mu
({\hat D}_\mu W_\nu -{\hat D}_\nu W_\mu )
- i g W_\mu ({\hat D}_\mu W_\nu - {\hat D}_\nu W_\mu )^*, \nn\\
&\hat{D}_\mu(\hat{D}_\mu W_\nu- \hat{D}_\nu W_\mu)=ig W_\mu
({\cal F}_{\mu\nu} +W_{\mu\nu}).
\end{eqnarray}
This shows that one can indeed Abelianize the non-Abelian theory
with our decomposition.
{\it An important point of the Abelian formalism is that, in addition to the
local dynamical degrees $W^1_\mu, ~W^2_\mu$, and $A_\mu$
of Yang-Mills theory, it has
an extra magnetic potential $\tilde C_\mu$.
Furthermore the  the Abelian potential
${\cal A}_\mu$ which couples to $W_\mu$ is given by the sum of
the electric and magnetic potentials
$A_\mu+\tilde C_\mu$. Clearly $\tilde C_\mu$ represents the topological
degrees of the non-Abelian symmetry which does not show up in the
naive Abelianization that one obtains by fixing the gauge} \cite{cho1,cho2}.

An important feature of this Abelianization is that
it is gauge independent, because here we have never fixed
the gauge to obtain this Abelian formalism. So one might
ask how the non-Abelian gauge symmetry is realized in this Abelian
formalism. To discuss this let
\bea
&\vec \alpha = \alpha_1~\hn_1 + \alpha_2~\hn_2 + \theta~\hat n,
~~~~~\alpha = \dfrac{1}{\sqrt 2} (\alpha_1 + i ~\alpha_2), \nn\\
&\vec C_\mu = - \dfrac {1}{g} \hn \times \partial_\mu \hn
= - C^1_\mu \hn_1 - C^2_\mu \hn_2,
~~~~~C_\mu = \dfrac{1}{\sqrt 2} (C^1_\mu + i ~ C^2_\mu).
\eea
Then the Lagrangian (\ref{eq:Abelian}) is invariant not only under
the active gauge transformation (\ref{gt1}) described by
\bea \label{eq:active}
&\delta A_\mu = \dfrac{1}{g} \partial_\mu \theta -
i (C_\mu^* \alpha - C_\mu \alpha^*),
~~~&\delta \tilde C_\mu = - \delta A_\mu,~~~\delta W_\mu = 0,
\eea
but also under the following passive gauge transformation
described by
\bea \label{eq:passive}
&\delta A_\mu = \dfrac{1}{g} \partial_\mu \theta -
i (W_\mu^* \alpha - W_\mu \alpha^*), ~~~&\delta \tilde C_\mu = 0,
~~~\delta W_\mu = \oneg \hD_\mu \alpha - i \theta W_\mu.
\eea
Clearly this passive gauge transformation assures the desired
non-Abelian gauge symmetry for the Abelian formalism.
This tells that the Abelian theory not only retains
the original gauge symmetry, but actually has an enlarged (both the
active and passive) gauge symmetries.
But we emphasize that this is not the ``naive'' Abelianization
of Yang-Mills theory which one obtains by fixing the gauge.
Our Abelianization is a gauge-independent Abelianization.
Besides, here the Abelian gauge
group is $U(1)_e \otimes U(1)_m$, so that
the theory becomes a dual gauge theory \cite{cho1,cho2}. This is
evident from (\ref{eq:active}) and (\ref{eq:passive}).

With this we can now obtain the gauge invariant Abelianization of
Georgi-Glashow model. From (\ref{eq:Tri}) and (\ref{chodecom}) we have
\begin{eqnarray}
\label{eq:lagran}
&{\cal L}_{GG} = -\dfrac{1}{2} ({\hat D}_\mu \vec \Phi)^2
-\dfrac{g^2}{2} (\vec W_\mu \times \vec \Phi)^2
-\dfrac{\lambda}{4}\left(\vec \Phi^2 -\frac{\mu^2}{\lambda}\right)^2 \nn\\
&- \dfrac{1}{4} (\D_\mu\W_\nu-\D_\nu\W_\mu)^2
- \dfrac{1}{4} (\hat F_{\mu\nu} + g \vec W_\mu \times \vec W_\nu)^2 \nn\\
&= -\dfrac{1}{2} (\partial_\mu \rho)^2
- g^2 {\rho}^2 W_\mu^*W_\mu
-\dfrac{\lambda}{4}\left(\rho^2 -\frac{\mu^2}{\lambda}\right)^2
-\dfrac{1}{2} (\D_\mu W_\nu-\D_\nu W_\mu)^2 \nn\\
&- \dfrac{1}{4} {\cal F}_{\mu\nu}^2
+ ig {\cal F}_{\mu\nu} W_\mu^*W_\nu
+ \dfrac{g^2}{4}(W_\mu^* W_\nu - W_\nu^* W_\mu)^2.
\end{eqnarray}
Now, the spherically symmetric ansatz of the Julia-Zee dyon
\begin{eqnarray}
\label{eq:Zee}
\vec \Phi&=&\rho(r)\hat{n} \nonumber,~~~~~\hn= \hat r, \\
\vec A_\mu&=&\frac{1}{g}A(r)\partial_\mu t\,\,\hat{n}
  + \frac{1}{g}(f(r)-1)\hat{n}\times\partial_\mu \hat{n},
\end{eqnarray}
can be written in this Abelian formalism as
\begin{eqnarray}
\rho &=& \rho(r) \nn\\
W_{\mu} &=& \frac{i}{g}\frac{f(r)}{\sqrt2}e^{i\varphi}
      (\partial_\mu \theta +i \sin\theta \partial_\mu \varphi), \nn\\
{\cal A}_{\mu} &=& \frac{1}{g}A(r) \partial_{\mu}t +
  \frac{1}{g}(1-\cos\theta) \partial_{\mu} \varphi.
\end{eqnarray}
With the ansatz one has the following equation of motion
\begin{eqnarray}
\label{eq:JuliaZee}
&&\ddot{f}- \frac{f^2-1}{r^2}f=\left(g^2\rho^2-A^2\right)f, \nonumber\\
&&\ddot{\rho}+\frac{2}{r}\dot{\rho} - 2\frac{f^2}{r^2}\rho =
     \lambda \left(\rho^2 - \frac{\mu^2}{\lambda} \right)\rho , \\
&&\ddot{A}+\frac{2}{r}\dot{A} -2\frac{f^2}{r^2} A=0. \nonumber
\end{eqnarray}
With the boundary condition
\bea
&f(0)=1,~~~~~A(0)=0,~~~~~\rho(0)=0, \nn\\
&f(\infty)=0,~~~~~A(\infty)=A_0,~~~~~\rho(\infty)=\rho_0,
\eea
one can integrate (\ref{eq:JuliaZee}) and obtain the Julia-Zee dyon.
Again it must be clear from (\ref{eq:JuliaZee}) that,
for a given magnetic charge, there
are always two dyons with opposite electric charges. 
Moreover, for the monopole solution with $A=0$,
the equation reduces to the following
Bogomol'nyi-Prasad-Sommerfield equation
in the limit $\lambda=0$
\begin{eqnarray}
\dot{f} \pm e \rho f=0, \nonumber
\end{eqnarray}
\begin{eqnarray}
\dot{\rho}\pm \frac{1}{er^2}(f^2-1)=0,
\label{pseq}
\end{eqnarray}
which has the analytic solution \cite{Julia}
\begin{eqnarray}
f = \dfrac{e\rho_0 r}{\sinh(e\rho_0r)},
~~~~~\rho= \rho_0\coth(e\rho_0r)-\frac{1}{er}.
\end{eqnarray}
Notice that the boundary condition $A(0)=0$ and $f(0)=1$ is crucial
to make the $SU(2)$ potential $\vec A_\mu$ regular at the origin.
\vskip 1em

\setcounter{section}{4}
\setcounter{equation}{0}
\section*{IV. Finite Energy Electroweak Dyon}
\indent

The above analysis tells us two things. First,
the Julia-Zee dyon is nothing but an Abelian dyon
whose singularity at the origin is regularized by the charged
vector field $W_\mu$ and scalar field $\rho$.
Secondly, the Cho-Maison dyon in the unitary gauge
can also be viewed as an Abelian dyon whose singularity at the origin
is only partly regularized by the weak bosons.
Obviously this makes the two dyons very similar to each other.
This suggests that one could also try to
make the energy of the Cho-Maison solutions
finite by introducing additional interactions and/or charged vector fields.
In this section we will present two ways which
allow us to achieve this goal along this line,
and construct analytic electroweak monopole and dyon solutions
with finite energy.
\vskip 1em
\noindent{\bf A. Electromagnetic Regularization}
\vskip 1em

We first regularize the magnetic singularity
with a judicious choice of an extra
electromagnetic interaction of the charged vector field
with the Abelian monopole \cite{cho97,cho02}.
This regularization provides a most economic way
to make the energy of the Cho-Maison solution finite,
because here we could use the already existing $W$-boson without
introducing a new source.

For this we need to Abelianize Weinberg-Salam model first.
With
\bea\label{decom}
\phi = \frac{1}{\sqrt{2}}\rho \xi,
~~~~~\n = -\xi^{\dagger} \vec \tau ~\xi,~~~~~(\xi^{\dagger} \xi = 1),
\eea
we have
\begin{eqnarray}
&{\cal L} =-\dfrac{1}{2}{(\partial_{\mu}\rho)}^2
 - \dfrac{\rho^2}{2} {|{\cal D}_{\mu} \xi |}^2
-\dfrac{\lambda}{8}\left(\rho^2-\dfrac{2\mu^2}{\lambda}\right)^2
-\dfrac{1}{4}{\vec F}_{\mu\nu}^2 -\dfrac{1}{4} G_{\mu\nu}^2 \nn\\
&= \dfrac{1}{2}{(\partial_{\mu} \rho)}^2
- \dfrac{\rho^2}{2} \left({|D_{\mu}\xi|}^2
- {|\xi^{\dagger} D_{\mu}\xi|}^2\right)
+ \dfrac{\rho^2}{2}{\left(\xi^{\dagger} D_{\mu}\xi
      - i \dfrac{g'}{2}B_{\mu}\right)}{}^2
-\dfrac{\lambda}{8}\left(\rho^2-\dfrac{2\mu^2}{\lambda}\right)^2 \nn\\
& -\dfrac{1}{4}({\cal F}_{\mu\nu}+W_{\mu\nu})^2 -\dfrac{1}{4} G_{\mu\nu}^2
-\dfrac{1}{2}|\hat D_\mu W_\nu -\hat D_\nu W_\mu |^2 \nonumber \\
&= -\dfrac{1}{2}(\partial_\mu \rho)^2
-\dfrac{g^2}{4} \rho^2 W_\mu^* W_\mu
 -\frac{\rho^2}{8}(g{\cal A}_\mu - g'B_\mu)^2
-\dfrac{\lambda}{8}\Big(\rho^2-\dfrac{2\mu^2}{\lambda}\Big)^2 \nn\\
&-\dfrac{1}{4}({\cal F}_{\mu\nu} +W_{\mu\nu})^2 -\dfrac{1}{4} G_{\mu\nu}^2
-\dfrac{1}{2}|\hat D_\mu W_\nu -\hat D_\nu W_\mu |^2.
\label{lag2}
\end{eqnarray}
There are three points to be emphasized here.
First, the Lagrangian is explicitly invariant under the $U(1)$ gauge
transformation of the $\xi$ field. This means
that Weinberg-Salam model can also be
viewed as a gauged $CP^1$ model \cite{Cho0}. This of course is why
Weinberg-Salam model allows the Cho-Maison solutions.
Secondly, the charged vector field $W_\mu$ is nothing but
the $W$-boson. Indeed, with the identification of $W_\mu$ as the physical
$W$-boson, the Lagrangian becomes formally identical to
what we have in the unitary gauge. But there is an important difference.
Here we did not obtain the above Lagrangian by fixing the gauge.
As a result our Abelian gauge potential which couples to $W$-boson
is given by $A_\mu+\tilde C_\mu$,
and has a dual structure. Thirdly, in this gauge invariant
Abelianization the electromagnetic potential and $Z$-boson are given by
\bea
A_\mu^{\rm (em)} = \dfrac{1}{\sqrt{g^2+g'^2}} (g'{\cal A}_\mu + g B_\mu),
~~~~~Z_\mu = \dfrac{1}{\sqrt{g^2+g'^2}} (g {\cal A}_\mu - g' B_\mu),
\eea
so that, in terms of the physical fields, the Lagrangian
(\ref {lag2}) is expressed by
\bea
&{\cal L} = -\dfrac{1}{2}(\partial_\mu \rho)^2
-\dfrac{g^2}{4} \rho^2W_\mu^* W_\mu
 -\dfrac{g^2+g'^2}{8} \rho^2 Z_\mu^2
-\dfrac{\lambda}{8}\Big(\rho^2-\dfrac{2\mu^2}{\lambda}\Big)^2 \nn\\
&-\dfrac{1}{4} {F_{\mu\nu}^{\rm (em)}}^2 -\dfrac{1}{4} Z_{\mu\nu}^2
-\dfrac{1}{2}|(D_\mu^{\rm (em)} W_\nu - D_\nu^{\rm (em)} W_\mu)
+ ie \dfrac{g}{g'} (Z_\mu W_\nu - Z_\nu W_\mu)|^2 \nn\\
& +ie F_{\mu\nu}^{\rm (em)} W_\mu^* W_\nu
+ie \dfrac{g}{g'}  Z_{\mu\nu} W_\mu^* W_\nu
+ \dfrac{g^2}{4}(W_\mu^* W_\nu - W_\nu^* W_\mu)^2,
\label{lag3}
\eea
where $D_\mu^{\rm (em)}=\partial_\mu+ieA_\mu^{\rm (em)}$.

Already at this level the above Lagrangian provides us 
an important piece of information. In the absence of 
the electromagnetic interaction (i.e., with $A_\mu^{\rm (em)}
= W_\mu = 0$) the Lagrangian describes a spontaneously broken
$U(1)_Z$ gauge theory, 
\bea
{\cal L} = -\dfrac{1}{2}(\partial_\mu \rho)^2
-\dfrac{g^2+g'^2}{8} \rho^2 Z_\mu^2
-\dfrac{\lambda}{8}\Big(\rho^2-\dfrac{2\mu^2}{\lambda}\Big)^2 
-\dfrac{1}{4} Z_{\mu\nu}^2,
\eea
which is nothing but the Ginsburg-Landau theory of superconductivity.
Furthermore, here $M_H$ and $M_Z$ corresponds to the coherence length
(of the Higgs field) and the penetration length (of the magnetic
field made of $Z$-field). So, when $M_H > M_Z$ (or $M_H < M_Z$),
the theory describes a type II (or type I) superconductivity, 
which is well known to admit the Abrikosov-Nielsen-Olesen 
vortex solution. 
This confirms the existence of Nambu's string in Weinberg-Salam
model. What Nambu showed was that he could make the string 
finite by attaching the fractionally charged monopole
anti-monopole pair to this string \cite{nambu}.  

To regularize the Cho-Maison dyon
we now introduce an extra interaction
${\cal L}_1$ to (\ref{lag3}),
\begin{eqnarray}
{\cal L}_1=i\alpha e F_{\mu\nu}^{\rm (em)} W_\mu^* W_\nu
               +\frac{\beta}{4}g^2(W_\mu^*W_\nu-W_\nu^*W_\mu)^2,
\label{int1}
\end{eqnarray}
where $\alpha$ and $\beta$ are arbitrary constants.
Notice that the extra interaction still respects
the gauge invariance of the theory, because with the decomposition
(\ref{chodecom}) the extra interaction can be expressed 
in a gauge invariant form with the help of the 
gauge covariant multiplet $\vec W_\mu$.
With this additional interaction the Lagrangian (\ref{lag3})
is modified to
\begin{eqnarray}
\label{Lag1}
&\hat {\cal L} = {\cal L} + {\cal L}_1 \nn\\
&= -\dfrac{1}{2}(\partial_\mu \rho)^2
-\dfrac{g^2}{4} \rho^2W_\mu^* W_\mu
-\dfrac{\lambda}{8}\Big(\rho^2-\dfrac{2\mu^2}{\lambda}\Big)^2
-\dfrac{1}{4} {F_{\mu\nu}^{\rm (em)}}^2 \nn\\
&-\dfrac{1}{2}|(D_\mu^{\rm (em)} W_\nu - D_\nu^{\rm (em)} W_\mu)
+ie \dfrac{g}{g'}(Z_\mu W_\nu - Z_\nu W_\mu)|^2
-\dfrac{1}{4} Z_{\mu\nu}^2
-\dfrac{g^2+g'^2}{8} \rho^2 Z_\mu^2 \nn\\
& + ie \dfrac{g}{g'} Z_{\mu\nu} W_\mu^* W_\nu
+ie(1+\alpha) F_{\mu\nu}^{\rm (em)} W_\mu^* W_\nu
+ (1+\beta) \dfrac{g^2}{4}(W_\mu^* W_\nu - W_\nu^* W_\mu)^2,
\end{eqnarray}
so that with the ansatz (\ref{ansatz2}) the energy of dyon is given by
\begin{eqnarray}
\label{eng2}
\hat E &=&\hat E_0 +\hat E_1, \nn\\
\hat E_0 &=& \dfrac{2\pi}{g^2}\int\limits_0^\infty
\dfrac{dr}{r^2}\bigg\{
\dfrac{g^2}{g'^2}+1-2(1+\alpha)f^2+(1+\beta)f^4 \bigg\}, \nn\\
\hat E_1 &=& E_1.
\end{eqnarray}
Notice that with $\alpha=\beta=0$, $\hat E_0$ reduces to $E_0$ and becomes
infinite.
For the energy (\ref{eng2}) to be finite, the integrand of
$\hat E_0$ must be free from both $O(1/r^2)$
and $O(1/r)$ singularities at the origin.
This requires us to have
\begin{eqnarray}
\label{cond1}
1+\frac{g^2}{g'^2}-2(1+\alpha) f^2(0)+(1+\beta) f^4(0)=0,  \nonumber
\end{eqnarray}
\begin{eqnarray}
(1+\alpha)f(0)-(1+\beta) f^3(0) =0 .
\end{eqnarray}
Thus we arrive at the following condition
for a finite energy solution
\begin{eqnarray}
\label{cond3}
1+\beta = (1+\alpha)^2 \sin^2\theta_{\rm w}
= (1+\alpha)^2 \dfrac{e^2}{g^2},
~~~~~f(0)=\frac{1}{\sqrt{(1+\alpha)\sin^2\theta_{\rm w}} }.
\end{eqnarray}
But notice that, although obviously sufficient for
a finite energy solution, this condition does not guarantee
the smoothness of the gauge potentials at the origin.
One might try to impose the condition
$f(0)=1$, because in this case the $SU(2)$ potential $\vec A_\mu$
of the ansatz (\ref{ansatz1}) becomes regular everywhere including the origin.
Unfortunately this condition does not remove
the point singularity of $B_\mu$ at the origin.

The condition for an an analytic solution is given by \cite{cho02}
\bea
\label{cond4}
\alpha = 0,~~~~~1+\beta=\dfrac{e^2}{g^2},
~~~~~f(0)=\dfrac{1}{\sin \theta_{\rm w}}=\dfrac{g}{e}.
\eea
Notice that this amounts to changing the coupling strength
of the $W$-boson quartic self-interaction from $g^2/4$
to $e^2/4$. To derive this analyticity condition it is important to remember
that Weinberg-Salam model, just like Georgi-Glashow model,
can be viewed as a gauged $CP^1$ model \cite{Cho0}. This means that
it can also be expressed by a Higgs triplet. To see this
we introduce a Higgs triplet $\vec \Phi$ and an ``electromagnetic'' $SU(2)$
gauge potential $\mbox{\boldmath$A$}_\mu$ by
\bea
&\vec \Phi = -\rho \xi^{\dagger} \vec \tau ~\xi = \rho \hn,
~~~~~(\phi = \dfrac{1}{\sqrt{2}} \rho \xi), \nn\\
&\mbox{\boldmath$A$}_\mu = (A_\mu^{\rm (em)}
-\dfrac{2i}{g} \xi^{\dagger} \partial_\mu \xi) \hn
- \dfrac{1}{e} \hn \times \partial_\mu \hn + \vec W_\mu
= \mbox{\boldmath$\hat A$}_\mu + \vec W_\mu.
\label{emsu2}
\eea
With this the electroweak Lagrangian (\ref{Lag1})
with (\ref{cond3}) can be expressed by
\begin{eqnarray}
\label{Lag2}
&\hat {\cal L} = -\dfrac{1}{2}(\mbox{\boldmath$\hat D$}_\mu \vec \Phi)^2
- \dfrac{g^2}{8} (\vec W_\mu \times \vec \Phi)^2
-\dfrac{\lambda}{8}\bigg(\vec \Phi^2 -\dfrac{2\mu^2}{\lambda}\bigg)^2 \nn\\
&-\dfrac{1}{4}(\mbox{\boldmath$\hat F$}_{\mu\nu}
+ (1+\alpha)e \vec W_\mu \times \vec W_\nu)^2
- \dfrac{1}{4} \Big(\mbox{\boldmath$\hat D$}_{\mu} \vec W_\nu
- \mbox{\boldmath$\hat D$}_\nu \W_\mu
+ e \dfrac{g}{g'} \hat n \times (Z_\mu \vec W_\nu
- Z_{\nu} \vec W_\mu) \Big)^2 \nn\\
&-\dfrac{1}{4} Z_{\mu\nu}^2
-\dfrac{g^2+g'^2}{8} \vec \Phi^2 Z_\mu^2
-\dfrac{e}{2} \dfrac{g}{g'} Z_{\mu\nu}
\hat n \cdot (\vec W_\mu \times \vec W_\nu),
\end{eqnarray}
where $\mbox{\boldmath$\hat D$}_\mu= \partial_\mu
+ e\mbox{\boldmath$\hat A$}_\mu \times$.
Notice that the Lagrangian is explicitly gauge invariant, 
due to the fact that $\vec W_\mu$ is gauge covariant.
This reassures that the extra interaction (\ref{int1})
is indeed gauge invariant.
Although the Lagrangian (because of the appearence of $\hn=\Phi/|\Phi|$)
looks to contain a non-polynomial interaction, the problematic
non-polynomial interaction disappears 
when it is expressed in terms of the physical fields.

In this form the Lagrangian describes
a ``generalized'' Georgi-Glashow model,
which has extra interaction with the $Z$-boson.
In particular, in the absence of the $Z$-boson, the theory reduces to
an $SU(2)_{\rm em}$ gauge theory
\begin{eqnarray}
\label{Lag0}
&\hat {\cal L} \rightarrow
-\dfrac{1}{2}(\mbox{\boldmath$D$}_\mu \vec \Phi)^2
+\dfrac{1}{2}(e^2-\dfrac{g^2}{4}) (\vec W_\mu \times \vec \Phi)^2
-\dfrac{\lambda}{8}\Big(\vec \Phi^2-\dfrac{2\mu^2}{\lambda}\Big)^2
-\dfrac{1}{4} \mbox{\boldmath $F$}_{\mu\nu}^2 \nn\\
&- \alpha \dfrac{e}{2} \mbox{\boldmath$F$}_{\mu\nu} \cdot
(\vec W_\mu \times \vec W_\nu)
- \alpha^2 \dfrac{e^2}{4} (\vec W_\mu \times \vec W_\nu)^2.
\end{eqnarray}
Evidently, with (\ref{cond4}), the Lagrangian
becomes almost identical to (\ref{eq:Julia}).
The only difference is
that here we have the extra interaction
$(e^2-g^2/4)(\vec W_\mu \times \vec \Phi)^2/2$.
Furthermore, in this form the ansatz (\ref{ansatz2}) is written as
\bea
\label{ansatz0}
\Phi&=&\rho(r)\hat{n},~~~~~\hn= \hat r, \nn\\
\mbox{\boldmath $A$}_\mu &=& e\Big(\frac{1}{g^2}A(r)+\frac{1}{g'^2}B(r)\Big)
~\partial_{\mu}t ~\hat n
+\dfrac{1}{e}(\dfrac{e}{g}f(r)-1) ~\hat{n}
\times \partial_{\mu} \hat{n}, \nn\\
Z_\mu&=&\frac{e}{gg'}(A(r)-B(r))\partial_\mu{}t .
\eea
Comparing this with the Julia-Zee ansatz (\ref{eq:Zee})
we conclude that the ansatz (\ref{ansatz2})
becomes smooth everywhere when $A(0)/g^2+B(0)/g'^2=0$
and $f(0)=g/e$. In particular the monopole singularity
disappears when $f(0)=g/e$.
This gives us the analyticity condition (\ref{cond4}).

But we emphasize that, to have a finite energy solution,
the condition (\ref {cond3}) is enough. Indeed, viewing the
electroweak theory as an Abelian gauge theory described by
(\ref{lag3}), there seems no apparent reason why the ansatz (\ref{ansatz2})
should satisfy the analyticity condition (\ref{cond4}). For this reason
we will leave $\alpha$ (and $f(0)$) arbitrary
in the following, unless specified otherwise.

With (\ref {cond3}) the equation of motion is
given by
\begin{eqnarray}
\ddot f - \dfrac{(1+\alpha)}{r^2}\Big(\frac{f^2}{f^2(0)}-1 \Big) f
=\Big(\frac{g^2}{4}\rho^2-A^2 \Big) f ,\nonumber
\end{eqnarray}
\begin{eqnarray}
\ddot \rho+\frac{2}{r}\dot\rho
-\frac{f^2}{2r^2}\rho
=-\frac{1}{4}(A-B)^2 \rho +\dfrac{\lambda}{2} \Big(\rho^2
-\frac{2\mu^2}{\lambda}\Big)\rho ,\nonumber
\end{eqnarray}
\begin{eqnarray}
\ddot A +\frac{2}{r} \dot A
- \frac{2f^2}{r^2} A
=\frac{g^2}{4}(A-B)\rho^2 ,
\label{eqm3}
\end{eqnarray}
\begin{eqnarray}
\ddot B+\frac{2}{r}\dot B
=-\frac{g'^2}{4}(A-B) \rho^2 .\nonumber
\end{eqnarray}
One could integrate this with the boundary conditions
near the origin,
\begin{eqnarray}
\label{origin1}
f/f(0)    &\simeq& 1 + \alpha_1  r^{\delta_1}, \nonumber \\
\rho &\simeq& \beta_1 r^{\delta_2}, \nonumber \\
A    &\simeq& a_1 r^{\delta_3},\\
B    &\simeq& b_0 + b_1 r^{\delta_4} ,
          \nonumber
\end{eqnarray}
and the finite energy condition (\ref {infty}) near the infinity.
Inserting (\ref {origin1}) to the equation we have
\bea
\label{origin2}
&\delta_1 = \dfrac{1}{2}(1+\sqrt{8\alpha+9}),
~~~~~\delta_2 = \dfrac{1}{2}(\sqrt{1+2f^2(0)} -1), \nn\\
&\delta_3 = \dfrac{1}{2}(\sqrt{1+8f^2(0)} -1),
~~~~~\delta_4 = \sqrt{1+2f^2(0)} +1.
\eea
Notice that all four deltas are positive (as far as $\alpha>-1$),
so that the four functions are well behaved at the origin.
Furthermore, when $\alpha=0$ and $f(0)=1$,
this reduces to (\ref{origin}). Clearly the solution
describes a finite energy electroweak dyon, even though
the gauge potential of the ansatz (\ref{ansatz0}) has
a (harmless) mathematical singularity at the origin
when $\alpha \ne 0$.

\begin{figure}
\epsfysize=7cm
\centerline{\epsffile{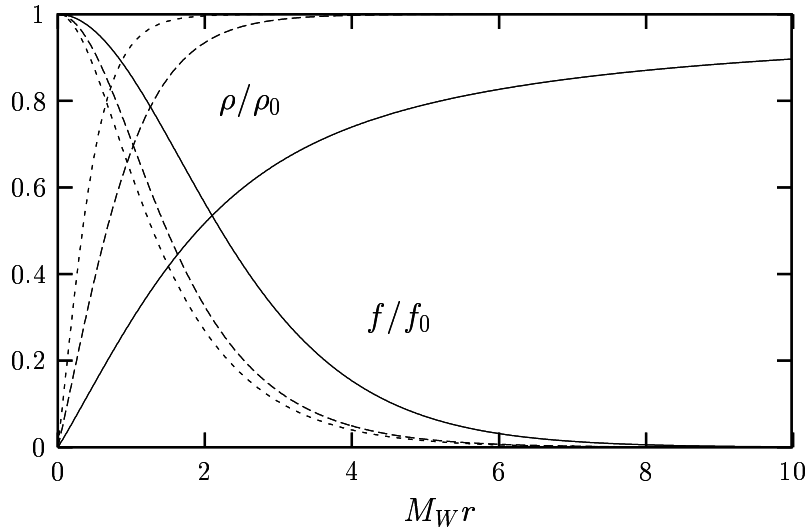}}
\caption{The finite energy electroweak monopole solution
obtained with different values of $\lambda/g^2=0$ (solid line),
$1/2$(dashed line), and $4$(dotted line).}
\label{fig2}
\end{figure}

\begin{figure}
\epsfysize=7cm
\centerline{\epsffile{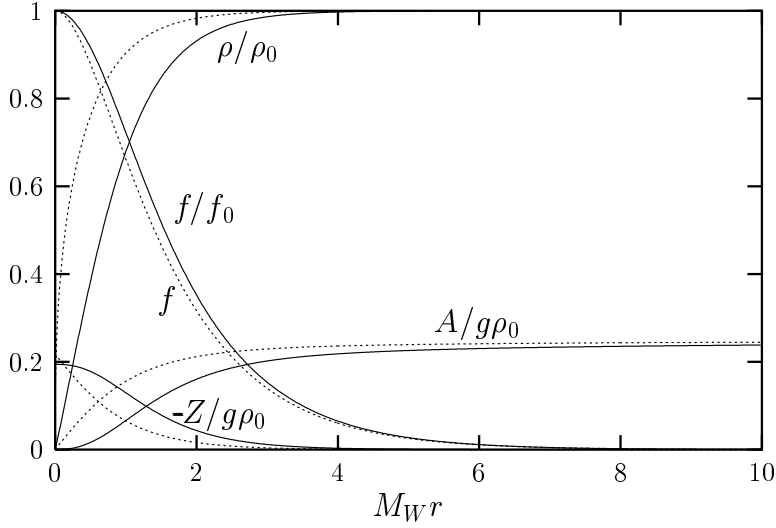}}
\caption{The electroweak dyon solution.
The solid line represents the finite energy dyon
and dotted line represents the Cho-Maison dyon, where we have chosen
$\lambda/g^2=1/2$ and $A_0=M_W/2$.}
\label{fig3}
\end{figure}

Now with the boundary condition
\bea
&&f(0)=g/e,~~~~~\rho(0)=0,~~~~~A(0)=0,~~~~~B(0)=b_0, \nn\\
&&f(\infty)=0,~~~~~\rho(\infty)=\rho_0,~~~~~A(\infty)=B(\infty)=A_0,
\label{bc1}
\eea
we can integrate (\ref{eqm3}) numerically.
Notice that, strictly speaking, the Coulomb potential
of the dyon retains a mathematical singularity at the origin
when $b_0\neq0$.
The results of the numerical integration for the monopole and dyon solution
are shown in Fig.\ref{fig2} and Fig.\ref{fig3}.
{\em It is really remarkable that the finite energy solutions look
almost identical to the Cho-Maison solutions,
even though they no longer have the magnetic singularity at the origin.
The reason for this similarity
must be clear. All that we need to have the
analytic monopole and dyon in the electroweak theory
is a simple modification
of the coupling strength of $W$-boson quartic self-interaction
from $g^2/4$ to $e^2/4$}.

Of course, with an arbitrary $\alpha$,
we can still integrate (\ref{eqm3}) and have
a finite energy solution. In this case
the gauge potential $\mbox{\boldmath $A$}_\mu$ in general has
a (harmless) mathematical singularity at the origin.
Even in this case, however,
the generic feature of the solutions remain the same.

Clearly the energy of the dyon must be of the order of $M_W$.
Indeed for the monopole the energy can be expressed as
\begin{eqnarray}
E=\frac{4\pi }{e^2} C(\alpha, \sin^2\theta_{\rm w},\lambda/g^2)M_W
\end{eqnarray}
where $C$ the dimensionless function of
$\alpha$, $ \sin^2\theta_{\rm w}$, and $\lambda/g^2$.
With $\alpha=0$ and experimental value $\sin^2\theta_{\rm w}$, $C$
becomes slowly varying function of $\lambda/g^2$ with $C=1.407$ for
$\lambda/g^2=1/2$.
This demonstrates that the finite energy solutions
are really nothing but the regularized
Cho-Maison solutions which have a mass of electroweak scale.

Notice that we can even have an explicitly analytic
monopole solution,
if we add an extra term ${\cal L}_2$ to the Lagrangian (\ref{Lag2})
\begin{eqnarray}
\label{mass}
{\cal L}_2=-\dfrac{1}{2}(e^2-\frac{g^2}{4}) 
(\vec W_\mu \times \vec \Phi)^2 =-(e^2-\frac{g^2}{4}) \rho^2W_\mu^* W_\mu.
\label{int2}
\end{eqnarray}
This amounts to changing the mass of the $W$-boson
from $g\rho_0/2$ to $e\rho_0$.
More precisely, with (\ref{cond4}) the extra term
effectively reduces the Lagrangian (\ref{Lag2})
to that of Georgi-Glashow model in the absence of
the $Z$-boson,
\begin{eqnarray}
\label{Lag3}
&\hat {\cal L} \rightarrow
-\dfrac{1}{2}(\mbox{\boldmath$D$}_\mu \vec \Phi)^2
-\dfrac{\lambda}{8}\Big(\vec \Phi^2-\dfrac{2\mu^2}{\lambda}\Big)^2
-\dfrac{1}{4} \mbox{\boldmath $F$}_{\mu\nu}^2.
\end{eqnarray}
Obviously this (with $A=B=0$) allows the well-known
Bogomol'nyi-Prasad-Sommerfield equation
in the limit $\lambda=0$ \cite{Julia}
\begin{eqnarray}
\dot{f}\pm e\rho f=0,
\nonumber
\end{eqnarray}
\begin{eqnarray}
\dot{\rho}\pm
\frac{1}{er^2}\Big(\frac{f^2}{f^2(0)}-1 \Big) =0.
\label{self2}
\end{eqnarray}
This has the analytic monopole solution
\begin{eqnarray}
f = f(0)\dfrac{e\rho_0 r}{\sinh(e\rho_0r)}
= \dfrac{g\rho_0 r}{\sinh(e\rho_0r)},
~~~~~\rho= \rho_0\coth(e\rho_0r)-\frac{1}{er},
\end{eqnarray}
whose energy is given by the Bogomol'nyi bound
\begin{eqnarray}
E=\dfrac{4\pi}{e^2}M'_{W}
= \dfrac{8\pi}{e^2}\sin \theta_{\rm w}M_{W}, ~~~~~(M'_{W}=e\rho_0).
\end{eqnarray}
But we emphasize that, even with this extra
term, the electroweak dyon becomes different from
the Prasad-Sommerfield dyon because it has a non-trivial dressing
of the $Z$-boson.

\vskip 1em
\noindent{\bf B. Embedding $SU(2)\times U(1)$ to $SU(2)\times SU(2)$}
\vskip 1em

As we have noticed the origin of the infinite energy of the Cho-Maison
solutions was the magnetic singularity of $U(1)_{\rm em}$. On the other hand
the ansatz (\ref{ansatz1}) also suggests that this singularity
really originates from the magnetic part of the hypercharge
$U(1)$ field $B_\mu$.
So one could try to  to obtain a finite energy monopole solution
by regularizing this hypercharge $U(1)$ singularity.
This could be done by introducing a hypercharged vector field
to the theory \cite{cho97}.
A simplest way to do this is, of course, to enlarge the hypercharge $U(1)$
and embed it to another $SU(2)$.

To construct the desired solutions we generalize the
Lagrangian (\ref{lag2}) by adding the following Lagrangian
\begin{eqnarray}
{\cal L}'&=&-\frac{1}{2}|\tilde D_\mu X_\nu-\tilde D_\nu X_\mu|^2
+ig' G_{\mu\nu}X_\mu^* X_\nu
+\frac{1}{4}g'^2(X_\mu^* X_\nu -X_\nu^* X_\mu)^2
\nonumber \\
&&-\frac{1}{2}(\partial_\mu\sigma)^2
-g'^2\sigma^2 X_\mu^* X_\mu
-\frac{\kappa}{4}\Big( \sigma^2-\frac{m^2}{\kappa}\Big)^2,
\label{lag4}
\end{eqnarray}
where
$X_\mu$ is a hypercharged vector field, $\sigma$ is a Higgs field, and
$\tilde D_\mu = \partial_\mu +ig' B_\mu$.
Notice that, if we introduce a hypercharge $SU(2)$ gauge field
$\vec B_\mu$ and a scalar triplet ${\vec \Phi}$ and
identify
\begin{eqnarray}
X_\mu =\frac{1}{\sqrt{2}}(B_\mu^1+i B_\mu^2)\nonumber,
~~~~~B_\mu = B_\mu^3 \nonumber,
~~~~~{\vec \Phi}= (0,0,\sigma),
\end{eqnarray}
the above Lagrangian becomes identical to
\begin{eqnarray}
{\cal L}'= -\frac{1}{2}(\tilde D_\mu {\vec \Phi})^2
-\frac{\kappa}{4}\Big({\vec \Phi}^2-\frac{m^2}{\kappa}\Big)^2
-\frac{1}{4} \vec G_{\mu\nu}^2,
\end{eqnarray}
as far as we interpret $B_\mu$ and $G_{\mu\nu}$ as the dual gauge field
of the hypercharge $U(1)$. This clearly shows that
Lagrangian (\ref{lag4}) is nothing but the embedding of the
hypercharge $U(1)$ to an $SU(2)$ Georgi-Glashow model.

From (\ref{lag2}) and (\ref{lag4})
one has the following equations of motion
\begin{eqnarray}
\partial_\mu(\partial_\mu \rho)=\frac{g^2}{2} W_\mu^* W_\mu \rho
+\frac{1}{4}(g {\cal A}_\mu - g'B_\mu)^2 \rho
+\dfrac{\lambda}{2}\Big(\rho^2-\frac{2\mu^2}{\lambda}\Big)\rho
,\nonumber
\end{eqnarray}
\begin{eqnarray}
\hat D_\mu (\hat D_\mu W_\nu -\hat D_\nu W_\mu)
=ig {\cal F}_{\mu\nu}W_\mu
-g^2 W_\mu (W_\nu W_\mu^* -W_\nu^* W_\mu)
+\frac{g^2}{4}\rho^2 W_\nu
,\nonumber
\end{eqnarray}
\begin{eqnarray}
\partial_\mu {\cal F}_{\mu\nu}&=&
\frac{g}{4}\rho^2(g {\cal A}_\nu-g'B_\nu)
+ig\Big( W_\mu^*(\hat D_\mu W_\nu -\hat D_\nu W_\mu)
        -(\hat D_\mu W_\nu -\hat D_\nu W_\mu)^*W_\mu\Big)
\nonumber \\
&&+ig \partial_\mu (W_\mu^* W_\nu -W_\nu^* W_\mu)
,\nonumber
\end{eqnarray}
\begin{eqnarray}
\partial_{\mu} G_{\mu\nu}&=&
\frac{g'}{4}\rho^2 (g'B_\nu-g {\cal A}_\nu)
  +ig'\Big( X_\mu^*(\tilde D_\mu X_\nu -\tilde D_\nu X_\mu)
             -(\tilde D_\mu X_\nu -\tilde D_\nu X_\mu)^*X_\mu\Big)
\nonumber \\
&&    + ig'\partial_\mu (X_\mu^* X_\nu-X_\nu^* X_\mu),
\nonumber
\end{eqnarray}
\begin{eqnarray}
\partial_\mu(\partial_\mu \sigma)=2g'^2 X_\mu^* X_\mu \sigma
+\kappa\Big( \sigma^2-\frac{m^2}{\kappa}\Big)\sigma ,\nonumber
\end{eqnarray}
\begin{eqnarray}
\tilde D_\mu (\tilde D_\mu X_\nu -\tilde D_\nu X_\mu)
  =ig'G_{\mu\nu} X_\mu
  -g'^2X_\mu (X_\mu^* X_\nu-X_\nu^* X_\mu)
  +g'^2\sigma^2 X_\nu
\label{eom5}
\end{eqnarray}
Now for a static spherically symmetric ansatz
we choose (\ref{ansatz1}) and assume
\begin{eqnarray}
\sigma &=&\sigma(r), \nonumber \\
X_\mu &=&\frac{i}{g'}\frac{h(r)}{\sqrt{2}}e^{i\varphi} (\partial_\mu \theta
+i\sin\theta\partial_\mu \varphi).
\label{ansatz3}
\end{eqnarray}
With the spherically symmetric  ansatz
(\ref{eom5}) is reduced to
\begin{eqnarray}
\ddot{f} - \frac{f^2-1}{r^2}f =
              \Big(\frac{g^2}{4}\rho^2 - A^2\Big)f, \nonumber
\end{eqnarray}
\begin{eqnarray}
\ddot{\rho} + \frac{2}{r} \dot{\rho} - \frac{f^2}{2r^2}\rho
  =- \frac{1}{4}(A-B)^2\rho + \dfrac{\lambda}{2}\Big(\rho^2 -
   \frac{2\mu^2}{\lambda}\Big)\rho \nonumber,
\end{eqnarray}
\begin{eqnarray}
\ddot{A} + \frac{2}{r}\dot{A} -\frac{2f^2}{r^2}A = \frac{g^2}{4}
   \rho^2(A-B), \label{eq:Spher}
\end{eqnarray}
\begin{eqnarray}
\ddot h -\frac{h^2-1}{r^2} h =(g'^2\sigma^2-B^2) h ,
\end{eqnarray}
\begin{eqnarray}
\ddot\sigma +\frac{2}{r}\dot\sigma -\frac{2h^2}{r^2} \sigma
= \kappa\Big(\sigma^2-\frac{m^2}{\kappa}\Big)\sigma , \nonumber
\end{eqnarray}
\begin{eqnarray}
\ddot{B} + \frac{2}{r} \dot{B}- \frac{2h^2}{r^2} B
 =  \frac{g'^2}{4} \rho^2 (B-A). \nonumber
\label{eom4}
\end{eqnarray}
Furthermore, the energy of the above
configuration is given by
\begin{eqnarray}
E=E_W +E_X,
\end{eqnarray}
\begin{eqnarray}
&E_{W}= \dfrac{4\pi}{g^2}\int\limits_0^\infty dr
\bigg\{(\dot f)^2 +\dfrac{(f^2-1)^2}{2r^2} +\dfrac{1}{2}(r\dot A)^2
+f^2A^2   \nonumber \\
&+\dfrac{g^2}{2}(r\dot\rho)^2 + \dfrac{g^2}{4} f^2\rho^2
+\dfrac{g^2r^2}{8}(A-B)^2\rho^2
+\dfrac{\lambda g^2r^2}{8}\Big(\rho^2-\frac{2\mu^2}{\lambda}\Big)^2
\bigg\}\nonumber \\
&=\dfrac{4\pi}{g^2}C_1(\lambda/g^2) M_W ,
\nonumber
\end{eqnarray}
\begin{eqnarray}
&E_X=\dfrac{4\pi}{g'^2}\int\limits_0^\infty dr\bigg\{(\dot h)^2
+\dfrac{(h^2-1)^2}{2r^2}
+\dfrac{1}{2}(r\dot B)^2 +h^2B^2 \nonumber \\
&+\dfrac{g'^2}{2}(r\dot\sigma)^2
+g'^2 h^2\sigma^2 +\dfrac{\kappa g'^2r^2}{4}(\sigma^2-\sigma_0^2)^2
\bigg\}\nonumber \\
&=\dfrac{4\pi }{g'^2} C_2(\kappa/g'^2)M_X ,
\nonumber
\end{eqnarray}
where $M_X=g'\sigma_0=g'\sqrt{m^2/\kappa}$.
The boundary conditions for a regular
field configuration can be chosen as
\begin{eqnarray}
&f(0)=h(0)=1,~~ A(0)=B(0)=\rho(0)=\sigma(0)=0, \nonumber\\
&f(\infty)=h(\infty)=0,~A(\infty)=A_0,~B(\infty)=B_0,
~\rho(\infty)=\rho_0,~\sigma(\infty)=\sigma_0.
\label{bound3}
\end{eqnarray}
Notice that this guarantees the analyticity of
the solution everywhere, including the origin.

\begin{figure}
\epsfysize=7cm
\centerline{\epsffile{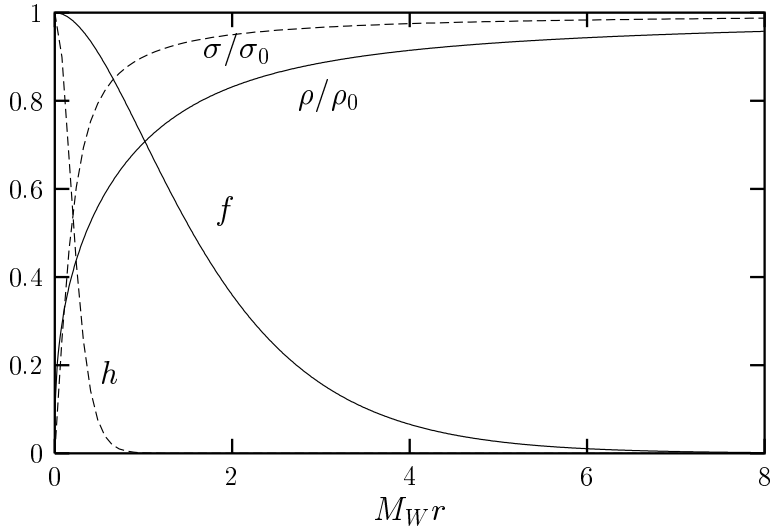}}
\caption{The $SU(2)\times SU(2)$ monopole solution, where
the dashed line represents hypercharge part which
describes Bogomol'nyi-Prasad-Sommerfield solution.}
\label{fig5}
\end{figure}

With the boundary condition (\ref{bound3}) one may try to find the desired
solution. From the physical point of view one could assume $M_X \gg M_W$,
where $M_X$ is an intermediate scale which lies somewhere between
the grand unification scale and the electroweak scale.
Now, let $A=B=0$ for simplicity. Then (\ref{eom4}) decouples to describes
two independent systems so that the monopole solution has two
cores, the one with the size $O(1/M_W)$ and the other with
the size $O(1/M_X)$. With $M_X=10M_W$
we obtain the
solution shown in Fig.\ref{fig5}
in the limit $\lambda=\kappa=0$.
In this limit we find
$C_1=1.946$ and $C_2=1$ so that the  energy of the solution
is given by
\begin{eqnarray}
E=\frac{4\pi}{e^2}\Big( \cos^2\theta_{\rm w}
          +0.195\sin^2\theta_{\rm w}\Big)M_X.
\end{eqnarray}
Clearly the solution describes
the Cho-Maison monopole whose singularity is regularized by
a Prasad-Sommerfield monopole of the size $O(1/M_X)$.

It must be emphasized that, even though the energy of the monopole
is fixed by the intermediate scale, the size
of the monopole is fixed by the electroweak scale. Furthermore from the
outside the monopole looks exactly the same as the Cho-Maison
monopole. Only the inner core is regularized by the hypercharged vector field.
This tells that the monopole should be interpreted
as an electroweak monopole.

\setcounter{section}{6}
\section*{V. Conclusions}
\indent

In this paper we have discussed two ways to regularize the Cho-Maison
monopole and dyon solutions of the Weinberg-Salam model,
and explicitly constructed
genuine finite energy electroweak monopole and dyon solutions
which are analytic everywhere including the origin.
The finite energy solutions are obtained with a simple modification
of the interaction of the $W$-boson or with the embedding of
the hypercharge $U(1)$ to a compact $SU(2)$.
It has generally been believed that the finite
energy monopole must exist only at the grand unification
scale~\cite{Dokos}. But our result tells that this belief is
unfounded, and endorses the existence of a totally new class of electroweak
monopole whose mass is much smaller than the monopoles of the
grand unification.
Obviously the electroweak monopoles are topological solitons which
must be stable.

Strictly speaking the finite energy solutions are not the solutions of the
Weinberg-Salam model, because their existence requires
a modification or generalization of the
model. But from the physical point of view there is no doubt that they should
be interpreted as the electroweak monopole and dyon, because
they are really nothing but the regularized Cho-Maison solutions whose size is
fixed at the electroweak scale. In spite of the fact that the Cho-Maison
solutions are obviously the solutions of the Weinberg-Salam model
one could try to object them as the electroweak dyons
under the presumption that the
Cho-Maison solutions could be regularized only at the grand unification scale.
Our work shows that
this objection is groundless,
and assures that
it is not necessary for us to go to the
grand unification scale to make the energy of the Cho-Maison solutions finite.
This really reinforces the Cho-Maison dyons as the
electroweak dyons which must be taken seriously.
Certainly the existence of the finite energy
electroweak monopoles should have important
physical implications~\cite{Preskill}.

We close with the following remarks:

\noindent{1)} A most important aspect of our result is that,
unlike the original Dirac monopole,
the magnetic charge of the electroweak monopoles must satisfy
the Schwinger quantization condition $q_m=4\pi n/e $.
Since the Weinberg-Salam model has an unbroken $U(1)_{\rm em}$,
one might try to embed the original Dirac monopole with the charge
$q_m =2\pi/e$ to it and obtain the monopole as a classical solution of the
Weinberg-Salam model. However, we emphasize that
this is possible strictly within the electrodynamics.
The electroweak unification simply forbids such an embedding.
So within the framework of the electroweak unification
the unit of the magnetic charge must be $4\pi/e$, not $2\pi/e$.
The existence of a monopole with $q_m=2\pi/e$ is simply not compatible with
the Weinberg-Salam model. This point has never been
well-appreciated before.

\noindent{2)} It has generally been believed that the topological
aspects of Weinberg-Salam model and Georgi-Glashow model are
quite different, because the Weinberg-Salam model is based on
a Higgs doublet but the Georgi-Glashow model is based on
a Higgs triplet. Our analysis shows that this is not true.
Both of them can be viewed as a gauged $CP^1$ model
which have exactly the same topology $\pi_2(S^2)$. Furthermore,
in the absence of the $Z$-boson, even the dynamics becomes very similar.
In fact we have shown that, with a simple change
of the mass and quartic self-interaction of
the $W$-boson, the two theories in this case become identical in
the limit $\lambda=0$. Only the presence of
the $Z$-boson in Weinberg-Salam model makes
them qualitatively different. This point also has not been
well-appreciated.

\noindent{3)} The electromagnetic regularization of
the Abelian point monopole with the charged vector fields
by the interaction (\ref{int1}) is nothing new.
In fact it is this regularization which makes the
energy of the 't Hooft-Polyakov monopole finite.
Furthermore it is well-known that
the 't Hooft-Polyakov monopole
is the only analytic solution (with $\alpha=\beta=0$)
which one could obtain with this technique~\cite{klee}.
What we have shown in this paper is that the same
technique also works to regularize the Cho-Maison solutions, but
with $\alpha=0$ and $\beta= -g^2/(g^2+g'^2)$.

\noindent{4)} The introduction of the additional interactions (\ref{int1})
and (\ref{int2}) to the Lagrangian (\ref{lag1}) could spoil the
renormalizability of the Weinberg-Salam model (although this issue has to
be examined in more detail). How serious would this offense be, however,
is not clear at this moment. The existence of the monopole
makes the renormalizability difficult to enforce in
the electroweak theory. Here we simply notice that
the introduction of a non-renormalizable interaction
has been an acceptable practice
to study finite energy classical solutions.

\noindent{5)} The embedding of the electroweak $SU(2)\times U(1)$ to
a larger $SU(2)\times SU(2)$ could
naturally arise in the left-right symmetric grand unification
models, in particular in the $SO(10)$ grand unification, although
the embedding of the hypercharge $U(1)$ to a compact $SU(2)$
may turn out to be too simple to be realistic.
Independent of the details, however, our discussion suggests
that the electroweak monopoles at an intermediate scale
$M_X$ could be  possible in a realistic grand unification.

For a long time it has been asserted that the standard
electroweak theory of Weinberg and Salam has no topological 
properties of interest. Obvoiusly this assertion is not 
based on the facts. We hope that our paper will correct 
this misunderstanding once and for all.  

\section*{Acknowledgments}
\indent

The work is supported in part by Korea Research Foundation (Grant KRF-2001
-015-BP0085) and by the BK21 project of Ministry of Education.


\thebibliography{99}
\bibitem{Dirac} P.A.M. Dirac, Phys. Rev. {\bf 74}, 817 (1948).
\bibitem{Wu} T.T. Wu and C.N. Yang, in {\it Properties of Matter under Unusual
           Conditions}, edited by
           H. Mark and S. Fernbach (Interscience, New York) 1969;
           Phys. Rev. {\bf D12}, 3845 (1975).
\bibitem{cho80} Y.M. Cho, Phys. Rev. Lett. {\bf 44}, 1115 (1980); Phys. Lett.
{\bf B115}, 125 (1982).
\bibitem{Hooft} G. 't Hooft, Nucl. Phys. {\bf B79}, 276 (1974);\\
           A.M. Polyakov, JETP Lett. {\bf 20}, 194 (1974).
\bibitem{Julia} B. Julia and A. Zee, Phys. Rev. {\bf D11}, 2227 (1975);\\
             M.K. Prasad and C.M. Sommerfield, Phys. Rev. Lett.
         {\bf 35}, 760 (1975).
\bibitem{vach} T. Vachaspati and M. Barriola, Phys. Rev. Lett. {\bf 69},
1867 (1992); \\
M. Barriola, T. Vachaspati, and M. Bucher, Phys. Rev. {\bf D50}, 2819 (1994).
\bibitem{Cho0} Y.M. Cho and D. Maison, Phys. Lett. {\bf B391}, 360 (1997).
\bibitem{Yang1} Yisong Yang, Proc. Roy. Soc. {\bf A454}, 155 (1998).
\bibitem{Yang2} Yisong Yang, {\it Solitons in Field Theory and
Nonlinear Analysis} (Springer Monographs in Mathematics), p. 322
(Springer-Verlag) 2001.
\bibitem{nambu} Y. Nambu, Nucl. Phys. {\bf B130}, 505 (1977); \\
T. Vachaspati, Phys. Rev. Lett. {\bf 68}, 1977 (1992).
\bibitem{vacha} T. Vachaspati, Nucl. Phys. {\bf B439}, 79 (1995). 
\bibitem{cho97} Y.M. Cho and Kyungtae Kimm, hep-th/9705213.
\bibitem{cho02} Y.M. Cho, hep-th/0210298, submitted to Phys. Rev. Lett.
\bibitem{Forg} P. Forg\'acs and N.S. Manton, Commun. Math. Phys.
           {\bf 72}, 15 (1980).
\bibitem{Dashen} R.F. Dashen, B. Hasslacher, and A. Neveu, Phys. Rev.
            {\bf D10}, 4138 (1974); \\
N.S. Manton, Phys. Rev. {\bf D28}, 2019 (1983).  
\bibitem{Bais} F.A. Bais and R.J. Russell, Phys. Rev. {\bf D11}, 2692
                                (1975);\\
           Y.M. Cho and P.G.O. Freund, Phys. Rev. {\bf D12}, 1711 (1975);\\
           Y.M. Cho and D.H. Park, J. Math. Phys. {\bf 31}, 695 (1990);\\
           P. Breitenlohner, P. Forg\'acs, and D. Maison, Nucl. Phys. {\bf B383},
           357 (1992);\\
           Y.M. Cho, J.W. Kim, D.H. Park, and J.H. Yoon,
           Phys. Lett. {\bf B308}, 23 (1993).
\bibitem{klee} K. Lee and E. Weinberg, Phys. Rev. Lett, {\bf 73}, 1203
                (1994);\\
               C. Lee and P. Yi, Phys. Lett. {\bf B348}, 100 (1995).
\bibitem{cho1}Y.M. Cho, Phys. Rev. {\bf D21}, 1080 (1980);
J. Korean Phys. Soc. {\bf17}, 266 (1984);
Phys. Rev. {\bf D62}, 074009 (2000).
\bibitem{cho2}Y.M. Cho, Phys. Rev. Lett. {\bf 46}, 302 (1981);
Phys. Rev. {\bf D23}, 2415 (1981);\\ 
W.S. Bae, Y.M. Cho, and S.W. Kimm, Phys. Rev. {\bf D65}, 025005 (2002);\\
Y.M. Cho, H.W. Lee, and D.G. Pak, Phys. Lett. {\bf B 525}, 347 (2002);\\
Y.M. Cho and D.G. Pak, Phys. Rev. {\bf D65}, 074027 (2002).
\bibitem{fadd} L. Faddeev and A. Niemi, Phys. Rev. Lett.
{\bf 82}, 1624 (1999); Phys. Lett. {\bf B449}, 214 (1999).
\bibitem{shab}S. Shabanov, Phys. Lett. {\bf B458}, 322 (1999);
{\bf B463}, 263 (1999); \\
              H. Gies, Phys. Rev. {\bf D63}, 125023 (2001).
\bibitem{bpst}A. Belavin, A. Polyakov, A. Schwartz, and Y. Tyupkin,
Phys. Lett. {\bf B59}, 85 (1975); \\
              G. 't Hooft, Phys. Rev. Lett. {\bf 37}, 8 (1976).
\bibitem{cho79}Y.M. Cho, Phys. Lett. {\bf B81}, 25 (1979).
\bibitem{Dokos} C.P. Dokos and T.N. Tomaras, Phys. Rev. {\bf D21}, 2940 (1980).
\bibitem{Preskill} J. Preskill, Phys. Rev. Lett. {\bf 43}, 1365 (1979);\\
           C.G. Callan, Phys. Rev. {\bf D25}, 2141 (1982);\\
           V.A. Rubakov, Nucl. Phys. {\bf B203}, 311 (1982).
\end{document}